\documentclass[11pt,tightenlines,eqsecnum,floats,aps,amsmath,amssymb,nofootinbib,prd,shownopacs,floatfix]{revtex4-2}

\usepackage{graphicx}
\usepackage{epstopdf}
\usepackage{latexsym}
\usepackage{amssymb}
\usepackage{amsmath}
\usepackage{color}
\usepackage{mathrsfs}
\usepackage{xparse}
\usepackage{float}
\usepackage{mathtools}
\usepackage{multirow}

\begin{document}


  \renewcommand\arraystretch{2}
 \newcommand{\bq}{\begin{equation}}
 \newcommand{\eq}{\end{equation}}
 \newcommand{\bqn}{\begin{eqnarray}}
 \newcommand{\eqn}{\end{eqnarray}}
 \newcommand{\nb}{\nonumber}
 \newcommand{\cb}{\color{blue}}
    \newcommand{\cc}{\color{cyan}}
        \newcommand{\cm}{\color{magenta}}
\newcommand{\rc}{\rho^{\scriptscriptstyle{\mathrm{I}}}_c}
\newcommand{\rd}{\rho^{\scriptscriptstyle{\mathrm{II}}}_c}
\NewDocumentCommand{\evalat}{sO{\big}mm}{%
  \IfBooleanTF{#1}
   {\mleft. #3 \mright|_{#4}}
   {#3#2|_{#4}}%
}
\newcommand{\PRL}{Phys. Rev. Lett.}
\newcommand{\PL}{Phys. Lett.}
\newcommand{\PR}{Phys. Rev.}
\newcommand{\CQG}{Class. Quantum Grav.}
\newcommand{\parallelsum}{\mathbin{\!/\mkern-5mu/\!}}

\title{Relating dust reference models to conventional  systems in manifestly gauge invariant perturbation theory}


\author{Kristina Giesel$^{1}$}
\email{kristina.giesel@gravity.fau.de}
\author{Bao-Fei Li $^{2}$}
\email{baofeili1@lsu.edu}
\author{Parampreet Singh$^2$}
\email{psingh@lsu.edu}
\affiliation{$^{1}$ Institute for Quantum Gravity,  Department of Physics, FAU Erlangen-N\"urnberg, Staudtstr. 7, 91058 Erlangen, Germany\\
$^{2}$ Department of Physics and Astronomy, Louisiana State University, Baton Rouge, LA 70803, USA}

\begin{abstract}
Models with dust reference fields in relational formalism have proved useful in understanding the construction of gauge-invariant perturbation theory to arbitrary orders in the canonical framework. These reference fields modify the dynamical equations for perturbation equations. However, important questions remain open on the relation with conventional perturbation theories of inflaton coupled to gravity and of multi-fluid systems, and on understanding modifications in terms of physical degrees of freedom. These gaps are filled in this manuscript for
Brown-Kucha\v{r} and Gaussian dust models, both of which involve three scalar physical degrees of freedom. We establish a relationship of these models with conventional inflationary and multi-fluid system of inflation and ordinary dust by introducing a set of gauge invariant variables on the reduced phase space of the dust reference models. We find the modifications due to dust clocks to Bardeen equation in the longitudinal gauge and Mukhanov-Sasaki equation in the spatially-flat gauge, in terms of physical degrees of freedom.  This results in a closed system of equations for all the degrees of freedom needed to explore the evolution of the scalar perturbations. Our analysis shows
for the first time
that even for two-fluid systems, there is a
natural choice of the set of gauge invariant variables for each chosen gauge  which not only offers a direct physical interpretation but also results in simplifications to the dynamical equations.
\end{abstract}
\maketitle
\section{Introduction}
\label{Intro}
\renewcommand{\theequation}{1.\arabic{equation}}\setcounter{equation}{0}
From the early days  in canonical general relativity (GR) the construction of gauge invariant quantities, so called Dirac observables, has played a pivotal role \cite{komar1958,bergman1961,bergman2} and is instrumental to address the problem of time in canonical  GR  \cite{Kuchar:1991qf,Anderson:2010xm}. The relational formalism  \cite{rovelli,rovelli2,Vy1994,dittrich,dittrich2,thiemann2006,Pons:2009cz,Pons:2010ad} provides a framework in which such Dirac observables can be constructed once a set of reference fields has been chosen. These Dirac observables become the elementary variables in the reduced phase space and their dynamics is generated by a so called physical Hamiltonian that is itself a Dirac observable and non-vanishing on the constraint surface. The relational formalism has been successfully used in various settings to extract dynamics in GR \cite{dittrich2,dt2006,ghtw2010I,ghtw2010II,gt2015,Ali:2015ftw,Giesel:2017roz,Giesel:2018opa,gsw2019,Giesel:2020bht,Han:2020iwk}, scalar-tensor theories \cite{hgm2015}, classical and quantum spherical symmetric models \cite{gtt2010,Han:2020uhb}, loop quantum gravity (LQG) \cite{gt2007,dgkl2010,hp2011,hp2012,gv2016,gv2017,Ali:2018vmt,hl2019I,hl2019II} and quantum cosmological models (see for eg.  \cite{aps2006a,aps2006b,aps2006c,Kaup,Blyth,acs2010,closed1,open,ads,ds1,ds2,radiation,Mielczarek:2010rq,Amemiya:2009pj,Gryb:2018whn,Gielen:2020abd,Giesel:2020raf}).
An interesting avenue to understand the role of reference fields lies in the cosmological perturbation theory where 
Brown-Kucha\v{r} \cite{Brown:1994py} and Gaussian dust \cite{kt1991} models have been
analyzed for instance in  \cite{ghtw2010II,Giesel:2020bht,Han:2020iwk}. While these studies indicate that dynamical equations for cosmological perturbations are modified due to dust reference fields, such as in \cite{Giesel:2020bht} where effects of dust reference field have been investigated for Mukhanov-Sasaki equation in the spatially-flat gauge, there are two important gaps in the studies so far. The first of these is related to the lack of insights on the modifications arising from dust reference fields in terms of physical degrees of freedom with a clear physical interpretation.  The second issue deals with relating these modifications to the physical predictions and for a comparison with the analysis in the absence of reference fields, in particular to the conventional perturbation theory for inflaton coupled to gravity, and to a multi-fluid system of inflaton and non-relativistic matter which has the same degrees of freedom as the dust reference models.  For this, one must understand the way the
 elementary Dirac observables in the reduced phase space of the dust models are related (i) to the gauge invariant quantities usually chosen in conventional cosmological perturbation theory such as the Bardeen potential or the Mukhanov-Sasaki variable, and (ii) to the multi-fluid systems where reference fields are not dust clocks.

In this work we discuss these relations for the longitudinal and spatially flat gauge and show that one can construct a map on the reduced phase space of the dust models to a new set of Dirac observables that is chosen in such a way that the comparison with the conventional choice of variables is simple and straightforward. To obtain such a map that relates different choices of gauge invariant quantities also becomes necessary and important if one is interested in the numerical implementation of the evolution equations for the linear perturbations and its comparison with other approaches.
Note that both Brown-Kucha\v{r} and Gaussian dust  reference models have additional degrees of freedom to GR. In the conventional approach one considers an FLRW spacetime as the background which is sourced by an inflaton field, whereas in the above dust models we consider a two-fluid system, an inflaton along with dust reference field, which determine the background evolution. As a consequence, the scalar sector of the linear perturbations contains three independent degrees of freedom in the configuration space, while for the conventional approach for inflaton coupled to gravity there is only one degree of freedom encoded in the Bardeen potential or the Mukhanov-Sasaki variable in the longitudinal gauge or the spatially flat gauge respectively. Hence, for the dust models even if we construct Bardeen potential-like and Mukhanov-Sasaki-like variables in the reduced phase space their equations of motion contain a fingerprint of the two additional physical degrees of freedom in the scalar sector present in both dust models. Thus, to extract any predictions we need to consider their system of coupled differential equations. Therefore, a pertinent question is how these two additional physical degrees of freedom should be chosen in the reduced phase space of the dust models and whether there exists a choice that simplifies the resulting set of coupled differential equations.  In addition, a pertinent question arises to also compare the dust reference field models with the ones with the same number of degrees of freedom.  This corresponds to the system of an inflaton with non-relativistic matter in which one may choose geometrical degrees of freedom as reference fields. Another goal of our work is to also relate the dust reference field models with the latter system in the conventional perturbation theory. Finally, as emphasized first by Bardeen \cite{bardeen}, the choice of gauge-invariant variables have a natural physical interpretation with curvature perturbations only in their respective gauges which are longitudinal and spatially-flat gauges for Bardeen variables and Mukhanov-Sasaki variable respectively. It turns out that for reference fields as geometrical degrees of freedom a connection can be established between the choice of clocks, gauge-fixing conditions and the gauge-invariant variables \cite{Giesel:2018opa}.

Let us compare the goals of our work with the existing  literature. Former work \cite{ghtw2010II,Giesel:2020bht} focused on constructing a Bardeen potential-like and Mukhanov-Sasaki-like variables and then deriving the corresponding equations of motion for them in the relational formalism.
 As shown in \cite{Giesel:2020bht}, in the dust models the Mukhanov-Sasaki equation involve contributions from the two additional physical degrees of freedom, but the physical interpretation of these degrees of freedom was not analyzed in detail.  An exercise on similar lines was earlier carried out for Bardeen equations \cite{ghtw2010II}, but  the final equation  still involved the energy momentum tensor of the inflaton and if one further expresses this equation into one that in the conventional approach yields a closed differential equation for the Bardeen potential, there are also dust contributions  that were not  explicitly derived. In particular, in the former work in \cite{ghtw2010II,Giesel:2020bht} the main focus was lying on a comparison of the dust reference models to conventional systems which only involve gravity and the inflaton. In this context one usually restricts to compare the gauge invariant dynamics of those gauge invariant variables which are present in both systems and this explains why one only considered the Mukhanov-Sasaki and Bardeen equation respectively. Thus, for both the Mukhanov-Sasaki equation and the Bardeen equation, the physical interpretation of dust modifications has been lacking. In this work, in addition to constructing the Bardeen potential-like and Mukhanov-Sasaki-like variables, we  further choose appropriately  the remaining two physical degrees of freedom such that they have a transparent physical interpretation. In this process it is also important that the resulting coupled differential equations do not get too complicated  to obtain primordial power spectrum and for comparison with the conventional inflationary and multi-fluid models. Note that  the additional terms in the Mukhanov-Sasaki equation due to the dust reference fields look rather complicated if we consider the Mukhanov-Sasaki-like variable as well as two elementary Dirac observables in reduced phase space of the dust models \cite{Giesel:2020bht}. As we will show here, by a different choice for the two additional gauge invariant variables these dust contributions have a clear physical interpretation at the gauge invariant level. The choice of variables we will present below is more adapted to two-fluid systems in the background and hence simplifies in this sense the comparison with conventional linear cosmological perturbation theory.
Furthermore, a clear physical interpretation of the choice of the additional two gauge invariant degrees is crucial in order to also compare the dust reference models to conventional two-fluid models at the gauge invariant level. Moreover, we find that the physical interpretation of these additional gauge invariant degrees of freedom can be used to further understand  the relationship between different gauges and their respective geometric clocks in the conventional approach in which usually a subset  of the geometric degrees of freedom are used to construct gauge invariant versions of the remaining degrees of freedom.

 In this manuscript, we  analyze four cases resulting from two dust models and two choices of gauges.  The dust models are the Brown-Kucha\v r and the Gaussian dust model and the gauges are the longitudinal and the spatially-flat one. The paper is structured as follows. In section \ref{sec:review} we briefly summarize the properties of the reduced phase in the dust model and present the evolution equations of the elementary Dirac observables that have been already derived in \cite{Giesel:2020bht}. These results are then taken as the starting point for section \ref{sec:BardMSeqn}, where after a brief motivation of the present analysis, the map between the Dirac observables in the reduced phase space of the dust models and the conventional choice of variables is constructed. This map is then used in subsection \ref{sec:BardeenMSlikeVar} to construct the Bardeen potential-like and Mukhanov-Sasaki-like variable for the dust models. In subsection \ref{sec:bardeeninbk} and \ref{sec:msinbk} the coupled system of evolution equations for the three physical scalar degrees of freedom in the Brown-Kucha\v{r} model is derived  for the longitudinal and spatially flat gauge respectively. Furthermore, the choice for the two additional variables next to the Bardeen potential-like and Mukhanov-Sasaki-like variable is discussed. The corresponding results for the Gaussian dust model are presented in subsection \ref{sec:BardMSinGD}, where the discussion is rather brief since many steps in the derivation of the two models are similar. A summary and conclusion of our results can be found in section \ref{sec:Concl}. The Table I summarizes the construction carried out in this manuscript and compares with the model of inflaton coupled to gravity as well as inflaton and non-relativistic matter as a multi-fluid system.

In this manuscript, we use the Planck units with $\hbar=c=1$ and keep the Newton's constant $G$ explicit.  We also set $\kappa=8\pi G$.

\section{Linear cosmological perturbation theory in the Brown-Kucha\v{r} and the Gaussian dust models}
\label{sec:review}
\renewcommand{\theequation}{2.\arabic{equation}}\setcounter{equation}{0}
As the starting point for this section we consider the Hamilton's equations of motion for the Dirac observables constructed in \cite{Giesel:2020bht}, where the  Gaussian and the Brown-Kucha\v{r} dust models were considered. In both of the models the reduced phase space contains 14 degrees of freedom encoded in the elementary Dirac observables. Of these 6 of them are geometric scalar degrees of freedom, 2 are vector and 4 are tensor degrees of freedom, while the remaining 2 are matter degrees of freedom. In the following we will restrict our discussion to the geometric scalar as well as the matter degrees of freedom. Following the notation of \cite{Giesel:2020bht} the corresponding Dirac observables of the scalar linear perturbations in the geometric sector are given by\footnote{Note that in \cite{Giesel:2020bht} the elementary Dirac observables were not denoted such as by ${\mathcal O}_E, {\mathcal O}_\psi,{\mathcal O}_{\delta\varphi}$ but just by their corresponding (capital)  letter, that is $E:={\mathcal O}_E, \psi:={\mathcal O}_\psi, \delta\Phi:={\mathcal O}_{\delta\varphi}$.}
\begin{equation}
{\mathcal O}_\psi=\frac{\delta^{ij}{\mathcal O}_{\delta q_{ij}}}{6A}, \quad \quad {\mathcal O}_{p_\psi}=\frac{\delta_{ij} {\mathcal O}_{\delta p^{ij}}}{6\mathcal P},\quad
{\mathcal O}_E=\frac{3}{4A}\Delta^{-2}\partial^{<i}\partial^{j>}{\mathcal O}_{\delta q_{ij}},\quad
{\mathcal O}_{p_E}=\frac{3}{4\mathcal P}\Delta^{-2}\partial_{<i}\partial_{j>}{\mathcal O}_{\delta p^{ij}}.
\end{equation}
Here ${\mathcal O}_{\delta q_{ij}}$ and ${\mathcal O}_{\delta p^{ij}}$ denote the Dirac observable associated with the linear perturbations of the ADM metric and their momenta, $A,{\mathcal P}$ are the Dirac observables of the square of the background scale factor and its momentum, $\Delta$ denotes the Euclidean Laplacian and $\partial_{<i}\partial_{j>}:=\partial_{(i}\partial_{j)}-\frac{1}{3}\delta_{ij}A\Delta$. Further, the Dirac observables for the matter contribution of the scalar field and its momenta are denoted by ${\mathcal O}_{\delta\varphi}$ and ${\mathcal O}_{\delta\pi_\varphi}$   respectively.

As presented in \cite{Giesel:2020bht} above Dirac observables can be constructed once some reference matter like the dust has been chosen. Since the dust reference fields come with four additional degrees of freedom coupled to gravity, the final number of physical degrees of freedom and hence independent gauge invariant quantities is increased by four compared to the system without dust that is usually considered in linearized cosmological perturbation theory. Once a specific kind of reference matter is chosen, a so called physical Hamiltonian that generates the dynamics on the reduced phase space can be obtained. In general, the physical Hamiltonian turns out to be different for different choices of reference matter. In \cite{Giesel:2020bht} the physical Hamiltonians for the Gaussian and Brown-Kucha\v{r} dust models were considered and the resulting Hamiltonian equations for the linear perturbations of the elementary Dirac observables in the reduced phase space were derived. In the case of the Brown-Kucha\v{r} dust model they are given by
\bqn
\label{bk1}
\dot {\mathcal O}_ \psi&=&2 H\left(\mathcal O_{p_\psi}-\frac{1}{2}\mathcal O_\psi\right)+H \mathcal O_\phi+\frac{\Delta \mathcal O_B}{3},\\
\label{bk2}
\dot {\mathcal O}_E&=&-4H\left(\mathcal O_E+\mathcal O_{p_E}\right)+\mathcal O_B,\\
\label{bk3}
\dot{\mathcal O}_{\delta \varphi}&=&\frac{\lambda_\varphi \bar{ \pi}_\varphi N}{A^{3/2}}\left(\mathcal O_\phi+\frac{\mathcal O_{\delta \pi_\varphi}}{\bar{\pi}_\varphi}-3 \mathcal O_\psi\right),\\
\label{bk4}
\dot {\mathcal O}_{p_\psi}&=&\frac{N^2}{6AH}\Delta\left(\mathcal O_\phi+\mathcal O_\psi-\frac{1}{3}\Delta \mathcal O_E\right)+\left(\frac{\kappa N^2p}{4H}-\frac{H}{2}\right)\left(\mathcal O_{p_\psi}-\frac{1}{2}\mathcal O_\psi\right)-\frac{\kappa N^2}{8 H}\delta T\nb\\&&-\left(\frac{1}{4}H+\frac{\kappa N^2 p}{8 H}\right)\mathcal O_\phi+\frac{\Delta \mathcal O_B}{6},\\
\label{bk5}
\dot {\mathcal O}_{p_E}&=&-\frac{N^2}{4AH}\left(\mathcal O_\phi+\mathcal O_\psi-\frac{1}{3}\Delta \mathcal O_E\right)+\left(\frac{5}{2}H+\frac{\kappa N^2}{4H}p\right)\left(\mathcal O_E+\mathcal O_{p_E}\right)-\mathcal O_B, \\
\label{bk6}
\dot{\mathcal O}_{\delta \pi_\varphi}&=&\frac{NA^{3/2}}{\lambda_\varphi}\left(-\frac{3V_{,\bar \varphi}}{2}\mathcal O_\psi-\frac{V_{,\bar \varphi}}{2}\mathcal O_\phi+\frac{1}{A}\Delta \mathcal O_{\delta\varphi}-\frac{V_{,\bar \varphi  \bar \varphi}}{2}\mathcal O_{\delta \varphi}\right)+\overline \pi_\varphi\Delta \mathcal O_B,
\eqn
here the background quantities, such as the scalar field $\bar \varphi$ and its momentum $\bar \pi_\varphi$, are labelled by an overbar, $H=\dot A/2A$ is the Hubble rate, $\mathcal O_\phi$ is the linear perturbation of the lapse function which vanishes identically in the Brown-Kucha\v r and Gaussian dust model,
$\mathcal O_B$  denotes s the scalar contribution to the perturbed shift vector. The latter vanishes in the Gaussian dust model and for the Brown-Kucha\v{r} model we have  $\mathcal O_B=\delta \mathcal E^{\mathrm{dust}}_{\parallelsum}/\left(\kappa A\overline C\right)$, where $\delta \mathcal E^{\mathrm{dust}}_{\parallelsum}$ denotes the perturbed momentum density of the dust. Besides,  $p=\lambda_\varphi \bar \pi^2_\varphi/2A^3-V/2\lambda_\varphi$ is the pressure of the scalar field and $V$ is twice the usual value of the potential of the scalar field, $\overline{C}$ denotes the geometric and scalar field contributions to the Hamiltonian background constraint. Further, we have $\overline  C=-\overline {\mathcal E}^\mathrm{dust}$ with $\overline {\mathcal E}^\mathrm{dust}$ denoting the background energy of the dust. A similar relation holds for the linear perturbations where $\delta C=-\delta{\mathcal E}^\mathrm{dust}$ and $\delta\hat{C}=-\delta{\mathcal E}^\mathrm{dust}_{\parallelsum}$, with $\delta\hat{C}$ denoting the scalar part of the spatial diffeomorphism constraint.  The linear perturbation of $T:={\mathcal O}_{q^{ij}}{\mathcal O}_{T_{ij}}$, where $T_{ij}$ denotes the spatial components of the energy momentum tensor of the scalar field, has the form
\bq
\delta T=-3\lambda_\varphi \frac{\bar \pi^2_\varphi}{A^3}\mathcal O_\psi+\frac{\lambda_\varphi \bar \pi_\varphi}{A^3}\mathcal O_{\delta \pi_\varphi}-\frac{V_{,\bar \varphi}}{2\lambda_\varphi}\mathcal O_{\delta \varphi}.
\eq
For the Gaussian dust model the first order Hamilton's equations are very similar and can be obtained directly from above equations by dropping all the terms involving  $\mathcal O_B$ since  $\mathcal O_B$ vanishes for the Gaussian dust model.
 Given the Hamilton's equations in (\ref{bk1})-(\ref{bk6}) for the elementary variables in the reduced phase space, it is straightforward to derive the equations of motion for three independent configuration variables $\mathcal O_E$, $\mathcal O_\psi$ and $\mathcal O_{\delta \varphi}$, which read
\bqn
\label{2ndOrd1}
\ddot {\mathcal O}_E+\frac{N^2}{3A}\Delta \mathcal O_E+\left(2H-\chi\right)\dot {\mathcal O}_E &=&\frac{N^2}{A}\mathcal O_\psi-\chi \mathcal O_B,\\
\label{2ndOrd2}
\ddot {\mathcal O}_{\delta \varphi}+\left(\frac{N^2}{2}V_{,\bar \varphi \bar \varphi}-\frac{N^2}{A}\Delta \right)\mathcal O_{\delta \varphi}+\left(3H-\frac{\dot N}{N}\right)\dot{\mathcal O}_{\delta \varphi}&=&-3\dot {\bar\varphi}\dot {\mathcal O}_\psi+\dot{\bar \varphi}\Delta \mathcal O_B,\\
\label{2ndOrd3}
\ddot {\mathcal O}_\psi+\left(2H-\chi\right)\dot {\mathcal O}_\psi-\frac{N^2}{3A}\Delta \mathcal O_\psi+\frac{N^2}{9A}\Delta^2 \mathcal O_E+\frac{\kappa \dot {\bar \varphi}}{4\lambda} \dot {\mathcal O}_{\delta \varphi}&=&\frac{\kappa N^2 V_{,\bar\varphi}}{8\lambda} \mathcal O_{\delta \varphi}-\frac{\chi}{3}\Delta \mathcal O_B,
\eqn
here $\chi$ is defined via
\bq
\chi=\frac{\dot H}{H}+\frac{H}{2}+\frac{\kappa N^2 p}{4H}.
\eq
Again, the corresponding equations of motion for the Gaussian dust model can be obtained from the set of equations in (\ref{2ndOrd1})-(\ref{2ndOrd3}) by dropping all terms that involve $\mathcal O_B$.

The equations of motion in (\ref{2ndOrd1})-(\ref{2ndOrd3}) describe the coupled differential equations of the independent Dirac observables and hence provide a gauge invariant evolution. These Dirac observables were constructed by choosing the dust fields
as reference fields and each individual quantity is manifestly gauge invariant, that is not only  up to linear order but also invariant under finite gauge transformations. As a consequence, also any combination of these variables as well as their temporal derivatives are manifestly gauge invariant quantities, that is something we will take advantage of in the following. If we compare the setup with what is usually done in the conventional linear perturbation theory, then even if one starts with a two fluid system in the background, one uses part of the geometric degrees of freedom as reference fields in prominent gauges such as the longitudinal or the spatially flat gauge. In this case one will also obtain 6 independent degrees of freedom in the reduced phase space of the scalar sector, but these are encoded in different gauge invariant variables. For the longitudinal gauge these are the Bardeen potential and its momentum and two independent gauge invariant variables and their momenta related to the dust fields. In the case of the spatially flat gauge one chooses the Mukhanov-Sasaki variable as well as two gauge invariant variables and their momenta related to the dust fields.
 Therefore, in order to compare our framework to the conventional choice of variables in the linear cosmological perturbation theory we adopt the strategy of considering specific combinations of  Dirac observables ${\mathcal O}_E,{\mathcal O}_\psi,{\mathcal O}_{\delta\varphi}$ and their momenta that are generalizations of  the Bardeen potential and the Mukhanov-Sasaki variable at the level of Dirac observables in presence of dust reference fields.  Recall that  in the conventional case  one only couples the inflaton to gravity and no further dust fields, then one obtains a closed second order differential equation for the Bardeen potential and the Mukhanov-Sasaki variable respectively. If we now add the additional dust fields in the conventional setup their gauge invariant extensions will also contribute to these differential equations and these contributions we will denote as dust contributions in the following. For instance in \cite{Giesel:2020bht} the Mukhanov-Sasaki equation for the Brown-Kucha\v{r} and the Gaussian dust model was derived and it has the following form
 \bq
\label{oldMSeqn}
\ddot Q+\frac{3}{2}\frac{\dot A}{A}\dot Q-\left(\frac{\Delta}{A}+\frac{3}{2}\frac{\dot A}{A}\frac{\dot Z}{Z}+\frac{\ddot Z}{Z}\right)Q=F^{\rm BK/G}_\mathrm{dust}.
\eq
with the Mukhanov-Sasaki-like variable in terms of the elementary Dirac observables given by
\bq
\label{4b8}
Q={\mathcal O}_{\delta\varphi}+  Z \left({\mathcal O}_\psi-\frac{\Delta}{3}{\mathcal O}_E\right),
\eq
where $Z=2\lambda_\varphi \frac{\overline\Pi_\Phi}{A\mathcal P}$ and $F^{\rm BK/G}_\mathrm{dust}$ is the additional term accounting for the contributions from the dust reference fields either in the Brown-Kucha\v{r} or Gaussian dust model. Its explicit form reads \cite{Giesel:2020bht} \footnote{Here the notation from \cite{Giesel:2020bht} are used as we cite the results from there. In particular, the definitions of  $\delta \mathcal E^{\mathrm{dust}}_{\parallelsum}$ and  $\delta \mathcal E^{\mathrm{dust}}$ in \cite{Giesel:2020bht} differ by a constant $\kappa$ from the definitions used in this manuscript in (\ref{dust1})-(\ref{dust2}). }
\bqn
\label{Fdustold}
&&F^{\rm BK/G}_\mathrm{dust}=\left(-\frac{3\kappa\lambda_\varphi \overline{\Pi}_\Phi}{2 A^3}+\frac{\kappa^2\lambda^2_\varphi\overline{\Pi}^3_\Phi}{2A^5\mathcal P^2}+\frac{\kappa^2\lambda_\varphi\overline{\Pi}_\Phi \overline{\mathcal E}^\mathrm{dust}}{2A^{7/2}\mathcal P^2}+\frac{\kappa V_{,\Phi}}{2A\mathcal P}\right)\delta \mathcal E^{\mathrm{dust}}_{\parallelsum}+\frac{\kappa \lambda_\varphi\overline{\Pi}_\Phi}{2A^{5/2}\mathcal P}\delta \mathcal E^{\mathrm{dust}}\nb\\
&&+\overline{\mathcal E}^{\mathrm{dust}}\Bigg[-\frac{\kappa\lambda_\varphi\overline{\Pi}_\Phi}{2A^{5/2}\mathcal P}\Delta {\mathcal O}_{E}-\frac{3\kappa Q}{4A^{3/2}}+ {\mathcal O}_{\delta\varphi}\left(\frac{3\kappa}{2A^{3/2}}-\frac{\kappa^2\lambda_\varphi \overline{\Pi}^2_\Phi}{2A^{7/2}\mathcal P^2}-\frac{\kappa\sqrt A V_{,\Phi}}{2\lambda_\varphi \mathcal P\overline{\Pi}_\Phi}-\frac{\kappa^2\overline{\mathcal E}^{\mathrm{dust}}}{2A^2\mathcal P^2}\right)\Bigg]~,\nb\\
\eqn
where $Q$ is understood as a function of the elementary Dirac observables ${\mathcal O}_E,{\mathcal O}_\psi,{\mathcal O}_{\delta\varphi}$. Although we know that by construction $F^{\rm BK/G}_\mathrm{dust}$  is manifestly gauge invariant, its structure as a function of ${\mathcal O}_E,{\mathcal O}_\psi,{\mathcal O}_{\delta\varphi}$ looks rather complicated. Furthermore, from the form obtained in \cite{Giesel:2020bht} it is not obvious that the dust contributions encoded in $F^{\rm BK/G}_\mathrm{dust}$ can be interpreted as gauge invariant quantities for the dust degrees of freedom, which would be pivotal for a comparison to the conventional choice of variables. In the next section we will show that this is indeed possible. We discuss a general strategy how for a given choice of   gauge specific combinations of the elementary Dirac observables in the reduced phase space can be choose as a set of variables that mimics the conventional choice of gauge invariant variables in cosmological perturbation theory. This then provides a map between the reduced phase space of the dust models and the one conventionally chosen for linear perturbations around a two fluid background cosmology. Furthermore, such a kind of map also allows to compare their corresponding gauge invariant equations of motion for the two choices of sets of gauge invariant variables.

\section{The Bardeen and Mukhanov-Sasaki equations in the Brown-Kucha\v r and the Gaussian dust models}
\label{sec:BardMSeqn}
\renewcommand{\theequation}{3.\arabic{equation}}\setcounter{equation}{0}
 For both models, namely the Brown-Kucha\v{r} and Gaussian dust models, as well as for both gauges the longitudinal as well as the spatially flat one we will follow the following strategy.  In former works \cite{ghtw2010II,Giesel:2020bht} these models and gauges were analyzed using a set of Dirac observables associated with metric and matter degrees of freedom. In order to compare the results to the conventional case where the dust is absent, one needs to identify a convenient set of three gauge invariant quantities in the scalar sector and their dynamics. In the former works such an identification has only been presented for one out of the three independent gauge invariant variables. Whereas for models that include an inflaton coupled to gravity, for each gauge only one independent gauge invariant variable exists. For the longitudinal gauge this corresponds to the Bardeen potential, where as for the spatially-flat gauge this gauge-invariant quantity is the Mukhanov-Sasaki variable. As expected the choice of gauge invariant variables is tightly connected to the chosen gauge. Moreover, in our case we investigate for each gauge two different dust models. Therefore, we expect for each chosen gauge to obtain a different set of three independent gauge invariant variables and since we consider two different dust models in general we expect for one chosen gauge different equations of motion for the gauge invariant variables depending on the chosen dust model. Because of this we will analyse the aforementioned four possible cases in this article. Further, we will show that for each gauge and each model, the choice of the set of gauge invariant variables is strongly connected with the choice of geometrical clocks in the conventional case \cite{gsw2019} where these geometrical clocks are gauged to vanish allowing to embed the conventional case into these dust models, where these geometrical clocks no longer vanish,  at the gauge invariant level.

In order to find a map that relates the set of independent physical degrees of freedom in the reduced phase space obtained by taking dust as reference fields and the conventional choice of variables we consider again the full phase space of all scalar degrees of freedom. This means in addition to $\mathcal{O}_E,\mathcal{O}_\psi,\mathcal{O}_{\delta\varphi}$ and their conjugate momenta we consider the Dirac observables $\mathcal{O}_\phi$ and $\mathcal{O}_B$ associated with the perturbed lapse function and scalar contribution to the perturbed shift vector as well as their conjugate momenta. Note that these are no independent degrees of freedom since on the reduced phase space these are functions of the independent Dirac observables. However, reconsidering them at this stage allows us to construct the desired map in a more systematic fashion. Next, we consider the way gauge invariant observables are constructed in linear cosmological perturbation theory and carry this over to our set of variables consisting of $\mathcal{O}_E,\mathcal{O}_\psi,\mathcal{O}_{\delta\varphi},\mathcal{O}_\phi$ and $\mathcal{O}_B$ and their momenta yielding
\bqn
\label{newvariables}
 \mathcal O^\mathrm{GI}_\phi&=&\mathcal O_\phi-\frac{1}{N}b_{,\tau},~~~~~~~~~~~  \mathcal O^\mathrm{GI}_{p_\phi}=\mathcal O_{p_\phi}, \nb\\
 \mathcal O^\mathrm{GI}_B&=&\mathcal O_B+\frac{N}{A}b-\hat b_{,\tau},~~~~~~~ \mathcal O^\mathrm{GI}_{p_B}=\mathcal O_{p_B},\nb \\
 \mathcal O^\mathrm{GI}_E&=&\mathcal O_E-\hat b, ~~~~~~~~~~~~~~~~~~ \mathcal O^\mathrm{GI}_{p_E}=\mathcal O_{p_E}+\frac{N b}{4AH}+\hat b,\nb\\
 \mathcal O^\mathrm{GI}_\psi&=&\mathcal O_\psi-\frac{H}{N}b-\frac{1}{3}\Delta \hat b,~~~~~\mathcal O ^\mathrm{GI}_{p_\psi}=\mathcal O_{p_\psi}+\left(\frac{H}{4N}+\frac{\kappa N}{8H}p\right)b-\frac{1}{6}\Delta\left(\frac{N}{AH}b+\hat b\right),\nb\\
 \mathcal O^\mathrm{GI}_{\delta  \varphi}&=&\mathcal O_{\delta \varphi}-\frac{\lambda_\varphi \bar \pi_\varphi}{A^{3/2}}b, ~~~~~~~~~\mathcal O^\mathrm{GI}_{\delta \pi_\varphi}=\mathcal O_{\delta \pi_\varphi}-\bar \pi_\varphi\Delta \hat b+\frac{A^{3/2}V_{,\bar \varphi}}{2\lambda_\varphi}b .
\eqn
Above the superscript $``GI"$ indicates that the new variables after the transformations are also gauge invariant quantities since   $b$ and $\hat b$, which are chosen to be  the gauge invariant analogues of the gauge descriptors in the conventional theory, are  functions of the elementary canonical variables in the reduced phase space.  With an appropriate choice of  $b$ and $\hat b$, one can find the observable analogues of the relevant  quantities in some particular gauges in the conventional theory as we  discuss in the next subsection. Note that for the two dust models considered in this work though $\mathcal{O}_\phi$  vanishes, however $\mathcal{O}^\mathrm{GI}_\phi$ is in general non-vanishing. The non-vanishing of the latter  will play an important role in our later discussion.

\subsection{The Bardeen potential and the  Mukhanov-Sasaki variable in the reduced phase space}
\label{sec:BardeenMSlikeVar}
In order to know the way the gauge invariant analogues of the gauge descriptors need to be chosen for a specific choice of gauge, we can apply the results obtained in \cite{Giesel:2017roz,Giesel:2018opa} on geometrical clocks in the relational formalism and carry them over to our situation here. Following the results from \cite{Giesel:2017roz,Giesel:2018opa} we choose for the longitudinal gauge  $b$ and $\hat b$ to be  the following functions on the reduced phase space
\bq
\label{lg}
b^L=-\frac{4AH}{N}\left(\mathcal O_E+\mathcal O_{p_E}\right),\quad \hat b^L=\mathcal O_E .
\eq
The analogues of the Bardeen potential and its momenta in the reduced phase space are then obtained as
\bqn
\label{bpotential}
\mathcal O^L_\psi&:=&\mathcal O_\psi+\frac{4AH^2}{N^2}\left(\mathcal O_E+\mathcal O_{p_E}\right)-\frac{1}{3}\Delta \mathcal O_E,\\
\mathcal O^L_{p_\psi}&:=&\mathcal O_{p_\psi}-\left(\frac{AH^2}{N^2}+\frac{\kappa A p}{2}\right)\left(\mathcal O_E+\mathcal O_{p_E}\right)+\frac{1}{2}\Delta \mathcal O_E+\frac{2}{3}\Delta \mathcal O_{p_E},
\eqn
where the superscript ``$L$" is used to denote physical observables in the reduced phase space in the longitudinal gauge. In the following, we simply refer to  $\mathcal O^L_\psi$ as the Bardeen potential in the longitudinal gauge. Similarly, one can find the remaining variables in the longitudinal gauge, which are
\bqn
\label{3a}
 \mathcal O_\phi^{L}&=&-\mathcal O_\psi^{L},\quad \mathcal O^{L}_{p_\phi}=\mathcal{O}_{p_\phi},\quad \mathcal O_E^L=0,\quad \mathcal{O}^L_{p_E}=0,\quad \mathcal{O}^L_{B}=0,\quad \mathcal{O}^L_{p_B}=\mathcal{O}_{p_B},
\eqn
where $\mathcal{O}_{p_B}$ and $\mathcal{O}_{p_\phi}$ are the  primary constraints. The corresponding new variables for the matter sector are given by
\bqn
\label{bardeenp}
\mathcal O^L_{\delta\varphi}&:=&\mathcal O_{\delta \varphi}+\frac{4\lambda_\varphi  \bar \pi_\varphi H}{NA^{1/2}}\left(\mathcal O_E+\mathcal O_{p_E}\right), \nb \\
\mathcal O^L_{\delta\pi_{\varphi}}&:=&\mathcal O_{\delta\pi_{\varphi}}-\bar \pi_\varphi \Delta \mathcal O_E-\frac{2H A^{5/2} V_{,\bar \varphi}}{\lambda_\varphi N}\left(\mathcal O_E+\mathcal O_{p_E}\right).
\eqn
Note that of course not all of these just constructed gauge invariant combinations are independent from each other, and some even vanish. However, considering all these variables and rewriting the Hamilton's equations of motion in terms of them, provides a systematic way to derive a second order differential equation of the Bardeen potential including some modifications due to the gauge invariant contributions from the dust energy and dust momentum density. This result then automatically leads to a convenient choice for the remaining two independent degrees of freedom in addition to the Bardeen potential in the reduced phase space.

Similarly if we are interested in the spatially flat gauge, then we aim at constructing the analogue of the Mukhanov-Sasaki variable in the reduced phase space by choosing
\bq
\label{sf}
b^S=\frac{N}{H}\left(\mathcal O_\psi-\frac{1}{3}\Delta \mathcal O_E\right),~~~~\hat b^S=\mathcal O_E,
\eq
which once plugged into (\ref{newvariables}) yields
\bqn
\label{ms}
\mathcal Q&:=&\mathcal O_{\delta \varphi}^{S}=\mathcal O_{\delta \varphi} -\frac{\lambda_\varphi \bar \pi_\varphi N}{H A^{3/2}}\left(\mathcal O_\psi-\frac{1}{3}\Delta \mathcal O_E\right), \nb\\
\label{3b1}
P_{\mathcal Q}&:=&\mathcal O_{\delta \pi_\varphi}^S=\mathcal O_{\delta \pi_\varphi}-\bar{\pi}_\varphi \Delta \mathcal O_E+\frac{NA^{3/2}V_{,\bar \varphi}}{2\lambda_\varphi H}\left(\mathcal O_\psi-\frac{1}{3}\Delta\mathcal O_E\right) .
\eqn
Above $\mathcal Q$ denotes the Mukhanov-Sasaki-like variable and $P_{\mathcal Q}$ its conjugate momentum in the reduced phase space, where capital $P$ is used to denote the conjugate momentum of the Dirac observables. The superscript ``$S$" is due to the fact that the Mukhanov-Sasaki variable has a natural physical interpretation in the spatially flat gauge in conventional perturbation theory (see for eg. \cite{Giesel:2018opa}). As in the case of the longitudinal gauge, one can find the remaining new variables for the choice of spatially flat gauge. These are given by
\bqn
\label{3b}
\mathcal O_\psi^{S}&:=&0,\quad \mathcal O_{p_\psi}^{S}:=\mathcal O_{p_\psi}+\left(\frac{1}{4}+\frac{\kappa N^2 p}{8H^2}-\frac{N^2\Delta}{6AH^2}\right)\left(\mathcal O_\psi-\frac{1}{3}\Delta \mathcal O_E\right)-\frac{1}{6}\Delta \mathcal O_E,
\nb \\
\mathcal O_\phi^{S}&:=&\mathcal O_\phi-\left(\frac{\dot N}{HN}-\frac{\dot H}{H^2}\right)\left(\mathcal O_\psi-\frac{1}{3} \Delta \mathcal O_E\right)-\frac{1}{H}\left(\dot {\mathcal O}_\psi-\frac{1}{3}\Delta \dot{\mathcal O}_E\right),\quad \mathcal O_{p_\phi}^{S}:=\mathcal{O}_{p_\phi},\nb\\ \mathcal O_B^{S}&:=&\frac{N^2}{AH}\left(\mathcal O_\psi-\frac{1}{3}\Delta \mathcal O_E\right)+4H\left(\mathcal O_E+\mathcal O_{p_E}\right),\quad  \mathcal O_{p_B}^{S}:=\mathcal{O}_{p_B},\quad\nb\\
\mathcal O_E^{S}&=&0,\quad \mathcal O_{p_E}^{S}=\frac{N^2}{4AH^2}\left(\mathcal O_\psi-\frac{1}{3}\Delta \mathcal O_E\right)+\mathcal O_E+\mathcal O_{p_E}.\quad
\eqn
Similar to the case in the longitudinal gauge, these elements of the set of variables are not independent, but the set provides a possibility to systematically identify the independent variables that are most convenient to compare our framework to the conventional one.

In the following, we derive the equations of motion of the  Bardeen potential and the Mukhanov-Sasaki variable in the reduced phase space of the  Brown-Kucha\v r  and the Gaussian dust models. As mentioned above,  there are three independent  scalar degrees of freedom in the linearized reduced phase space which are in general coupled to one another. Hence, neither the Bardeen potential nor the Mukhanov-Sasaki variable on its own can fully describe the evolution of the linear perturbations in the dust models.  It is necessary to find  two further independent variables in the reduced phase space that form together along with the Bardeen potential and the Mukhanov-Sasaki variable respectively a closed system of evolution equations.
Our guiding principle will be that by construction we know that these two additional independent variables are related to the gauge invariant extensions of the dust degrees of freedom. Once we have the equations of motions for the linear scalar perturbations in that form, we can easily identify the additional two independent variables. We explicitly show how to choose these two variables and moreover derive their  coupled system of differential equations they build together with the (generalized) Bardeen  and (generalized) Mukhanov-Sasaki equation respectively.

\subsection{The Bardeen equation in the Brown-Kucha\v r dust model}
\label{sec:bardeeninbk}
In this subsection, we  derive  the Bardeen equation  and the equations of motion of the other two variables which, coupled with the Bardeen potential, form a closed system for the evolution of the linear scalar perturbations in the Brown-Kucha\v r dust model. The starting point is the observation that the transformations in (\ref{newvariables}) are compatible with the equations of motion  (\ref{bk1})-(\ref{bk6}) for the elementary variables in the reduced phase space. That is to say,  the set of  new variables obtained from (\ref{newvariables}) also satisfy the  equations  (\ref{bk1})-(\ref{bk6}) once the old variables in these equations are promoted to  the new ones. It is easy to check that all the additional terms related with the gauge descriptors are cancelled exactly if one solves for the old variables in terms of the new ones and then substitute them back into   (\ref{bk1})-(\ref{bk6}). As a result,  the Hamilton's equations of motion for the Bardeen potential and its momentum are given by
\bqn
\label{3a1}
\dot {\mathcal O}_\psi^{L}&=&2H\left(\mathcal O_{p_\psi}^L-\mathcal O_\psi^L\right),\\
\label{3a2}
\dot {\mathcal O}_{p_\psi}^L&=&\left(-\frac{1}{2}H+\frac{\kappa N^2}{4H}p\right)\left(\mathcal O_{p_\psi}^L-\frac{1}{2}\mathcal O_\psi^L\right)-\frac{\kappa N^2 \delta T^L}{8 H}+\left(\frac{1}{4}H+\frac{\kappa N^2}{8 H}p\right)\mathcal O_\psi^L,
\eqn
where we have used  $\mathcal O_\phi^L=-\mathcal O_\psi^L$ and $\mathcal O^L_B=0$. Moreover,  the linear perturbation of the trace of the spatial components of the energy momentum tensor in the longitudinal gauge is given by
\bq
\label{3a3}
\delta  T^L=-3\lambda_\varphi \frac{\bar \pi^2_\varphi}{A^3}\mathcal O_\psi^L+\frac{\lambda_\varphi \bar \pi_\varphi}{A^3}\mathcal O^\mathrm{L}_{\delta \pi_\varphi}-\frac{V_{,\bar \varphi}}{2\lambda_\varphi}\mathcal O_{\delta  \varphi}^{L}.
\eq
Taking the time derivative of (\ref{3a1}) and with the help of (\ref{3a2}), it is straightforward to obtain the equation of motion for the Bardeen potential in the reduced phase space which turns out to be
\bq
\label{bardeen}
\ddot {\mathcal O}_\psi^L=\left(\frac{\dot N}{N}-4H\right)\dot {\mathcal O}_\psi^L+\frac{\kappa N^2 p}{2}{\mathcal O}_\psi^L-\frac{\kappa N^2}{4} \delta  T^{L}.
\eq
In order to  compare (\ref{bardeen}) with its counterpart in the conventional perturbation theory, one needs to express $\mathcal O^L_{\delta \pi_\varphi}$ and $\mathcal O_{\delta  \varphi}^{L}$ in $ \delta  T^{L}$ in terms of the Bardeen potential,  its time derivative and the contributions corresponding to the dust. This can be achieved by perturbing the total Hamiltonian and diffeomorphism constraints, which yields
\bqn
\label{dust1}
-\delta \mathcal E^\mathrm{dust}&=&\delta C = 4\sqrt A \Delta \left( \mathcal O_\psi-\frac{1}{3}\Delta \mathcal O_E\right)-\frac{3\sqrt A \mathcal P^2}{2}\left(\mathcal O_\psi+4\mathcal O_{p_\psi}\right)\nb\\&&+\kappa A^{3/2}\left(\frac{V_{,\bar \varphi}\mathcal O_{\delta \varphi}}{2\lambda_\varphi}+\frac{\lambda_\varphi \pi_\varphi \mathcal O_{\delta \pi_\varphi}}{A^3}-3p\mathcal O_\psi\right),\\
\label{dust2}
-\delta \mathcal E^\mathrm{dust}_{\parallelsum}&=&\delta\hat{C}=\kappa \pi_\varphi \mathcal O_{\delta\varphi}-4A\mathcal P\left(\frac{2}{3}\Delta \left(\mathcal O_E+\mathcal O_{p_E}\right)+\mathcal O_{p_\psi}-\frac{\mathcal O_\psi}{2}\right).
\eqn
Although the above formulae are in terms of the elementary Dirac observables, they can be  easily expressed  in terms of our new variables in the longitudinal gauge. The resulting equations are,
\bqn
\label{3a4}
-\delta  { \mathcal E}^\mathrm{dust}_L&=&\frac{-3\mathcal P^2\sqrt{A}}{2}\left(\mathcal O_\psi^L+4 \mathcal O_{p_\psi}^L\right)+4\sqrt A \Delta\mathcal O_\psi^L+3\kappa  \bar {\mathcal E}^\mathrm{dust}\mathcal O_\psi^L\nb\\&&+\kappa A^{3/2}\left(\frac{V_{,\bar \varphi}}{2\lambda_\varphi}\mathcal O_{\delta  \varphi}^L+\frac{\lambda_\varphi \bar \pi_\varphi}{A^3}\mathcal O_{\delta \pi_{\varphi}}^L-3p\mathcal O_\psi^L\right)=:\delta C_L,\\
\label{3a5}
-\delta  { \mathcal E}^\mathrm{dust}_{\parallelsum,L}&=&\kappa \pi_\varphi\mathcal O_{\delta \varphi}^L-4A\mathcal P\left(\mathcal O_{p_\psi}^L-\frac{\mathcal O_\psi^L}{2}\right)=:\delta\hat{C}_{\parallelsum,L} .
\eqn
Here we have introduced  $\delta  {\mathcal E}_L$ and $\delta  { \mathcal E}^\mathrm{dust}_{\parallelsum,L}$ which are the analogues of the gauge invariant perturbations of the dust energy and momentum density in the longitudinal gauge in a two fluid system. As can be seen in our case on the reduced dust space, these can be expressed completely in terms of the geometric and inflaton degrees of freedom. Their explicit form reads
\bqn
\label{3a6}
\delta { \mathcal E}^\mathrm{dust}_L&=&-\delta C_L=-\delta C +\kappa  \bar {C}\left(3\mathcal O_\psi^L+ \Delta \hat b^L\right),\\
\label{3a7}
\delta  { \mathcal E}^\mathrm{dust}_{\parallelsum,L}&=&-\delta\hat{C}_{\parallelsum,L}=-\delta \hat{C}+\kappa \bar {C} b^L,
\eqn
and $\bar {C}$ is the contribution of geometry and the inflaton to total Hamiltonian constraint of the background given by
\bq
\bar {C}=-\frac{3\mathcal P^2\sqrt A}{2\kappa}+\frac{\lambda_\varphi \bar\pi^2_\varphi}{2A^{3/2}}+\frac{V A^{3/2}}{2\lambda_\varphi}.
\eq
Let us compare this to the situation of a system involving gravity and the inflaton only where one uses the conventional linearized geometrical clocks. In this case $\delta C_L$ and $\delta\hat{C}_{\parallelsum,L}$ both vanish because then $\bar{C}$ and $\delta C$ are the total background and linearised Hamiltonian constraint. Moreover $b^L$ is the gauge fixing condition of the longitudinal gauge.

Now it is straightforward to solve for $\mathcal O_{\delta \varphi}^L$ and $\mathcal O_{\delta \pi_\varphi}^L$ from (\ref{3a4})-(\ref{3a5}), to obtain
\bqn
\label{3a8}
\mathcal O_{\delta \varphi}^L&=&\frac{2A\mathcal P}{\kappa \bar \pi_\varphi}\left(\mathcal O_\psi^L+\frac{\dot {\mathcal O}_\psi^L}{H}\right)+\frac{\delta \hat{C}_{\parallelsum,L}}{\kappa \bar \pi_\varphi },\\
\label{3a9}
\frac{\lambda_\varphi \bar \pi_\varphi}{A^3}\mathcal O_{\delta \pi_\varphi}^L&=&\frac{\delta C_{L}}{\kappa A^{3/2}}+\frac{3\mathcal P^2}{2\kappa A}\left(5 \mathcal O_\psi^L+\frac{2}{H}\dot {\mathcal O}_\psi^L\right)-\frac{4}{\kappa A} \Delta \mathcal O_\psi^L\nb\\&&+3\left(p+\frac{\bar{C}}{A^{3/2}}\right)\mathcal O_\psi^L-\frac{V_{,\bar \varphi}}{2\lambda_\varphi}\mathcal O_{\delta \varphi}^L,
\eqn
where we have used $\mathcal O_{p_\psi}^L=\mathcal O_\psi^L+\dot{\mathcal O}_\psi^L/(2H)$. Plugging (\ref{3a8})-(\ref{3a9}) into (\ref{bardeen}), the Bardeen equation finally takes its form as
\bqn
\label{3a10}
\ddot {\mathcal O}_\psi^L&-&\left(\frac{\dot N}{N}-7H-\frac{N^2V_{,\bar \varphi}}{\dot {\bar\varphi}}\right)\dot{\mathcal O}_\psi^L-\frac{N^2}{A}\Delta \mathcal O_\psi^L-\left(\frac{\kappa N^2 p}{2}-3H^2-\frac{N^2  V_{,\bar \varphi}}{\dot{\bar \varphi}}H\right)\mathcal O_\psi^L\nb\\&&~~~~~~~~~~~~~~~~~~~=\frac{N^2}{4A^{3/2}}\delta  { \mathcal E}^\mathrm{dust}_{L}-\frac{N^3 V_{,\bar \varphi}}{4A^{3/2}\dot{\bar \varphi}}\delta  { \mathcal E}^\mathrm{dust}_{\parallelsum,L}\nb\\&&~~~~~~~~~~~~~~~~~~~=-\frac{N^2}{4A^{3/2}}\delta  {C}_{L}+\frac{N^3 V_{,\bar \varphi}}{4A^{3/2}\dot{\bar \varphi}}\delta\hat{C}_{\parallelsum,L}
\equiv F^B_\mathrm{dust},
\eqn
where $ F^B_\mathrm{dust}$ stands for the dust contributions in the Bardeen equation. We emphasize that here  $ F^B_\mathrm{dust}$ is understood as an additional contribution to the Bardeen equation if one compares to the system of an inflaton minimally coupled to gravity.  
As compared with this counterpart system, the above Bardeen equation in the Brown-Kucha\v r dust model has two additional terms on the right hand side which are proportional to the perturbations of the dust energy density and dust momentum density respectively. Note that if one adds dust to the inflaton-gravity system in the conventional framework, then one obtains the same form of the above modified Bardeen equation albeit with variables which are gauge-invariant only till the linear order.  
In contrast, the  $ F^B_\mathrm{dust}$  term computed above using dust reference fields is a manifestly gauge invariant quantity. In absence of dust,  $\delta C_L$ and $\delta\hat{C}_{\parallelsum,L}$ vanish and so does  $ F^B_\mathrm{dust}$, one rediscovers the usual Bardeen equation in this limit.

Note that in our case  the Bardeen equation itself does not describe the evolution of  a closed system since the gauge invariant perturbations of the dust energy density and the dust momentum density also depend on other independent gauge invariant variables. Considering the form of $ F^B_\mathrm{dust}$, a convenient choice for the two further physical degrees of freedom would be the analogue of the gauge invariant extensions of the dust's energy and momentum densities, that is $\delta  { \mathcal E}^\mathrm{dust}_L$ and $\delta  { \mathcal E}^\mathrm{dust}_{\parallelsum,L}$. Given their explicit forms in (\ref{3a6}) and (\ref{3a7}) together with the fact that $\delta  {C}$ and $\delta\hat{C}$ are constants of motion in the Brown-Kucha\v{r} model, we can also choose $b_L$ and $\hat{b}_L$ as the two gauge invariant variables in addition to the Bardeen potential. In the conventional approach $b_L$ and $\hat{b}_L$ would not be gauge invariant, but for the dust reference models they are built from elementary manifestly gauge invariant Dirac observables and thus any function of them is again gauge invariant. For the reason that $\hat{b}_L=\mathcal{O}_E$, we work directly with $\mathcal{O}_E$ and choose in the case of the longitudinal gauge a set of gauge invariant quantities consisting of ($\mathcal O^L_\psi$, $b^L$, $\mathcal O_E$). With this choice we need to express $F^B_\mathrm{dust}$ in terms of these gauge invariant variables and moreover also derive the evolution equations for these to obtain the system of differential equations that describe the evolution of the linear perturbations in the Brown-Kucha\v{r} model. In particular, substituting  (\ref{3a6}) and (\ref{3a7})  into the expression of  $F^B_\mathrm{dust}$, we obtain
\bq
\label{dustinbardeen}
F^B_\mathrm{dust}=-\frac{N^2}{4A^{3/2}}\delta {C}+\frac{\kappa \bar{C} N^2}{4A^{3/2}}\left(3\mathcal O^L_\psi+\Delta \mathcal O_E\right)+\frac{N^3 V_{,\bar \varphi}}{4A^{3/2}\dot{\bar \varphi}} \left(\delta  \hat{C}+\kappa \overline{C}b^L \right).
\eq
The Hamilton's equation of $\mathcal O_E$ can be read from (\ref{bk2}), which, after introducing a new variable\footnote{Similar to the notation for the conjugate momentum of the Mukhanov-Sasaki variable in (\ref{3b1}), we use capital $P$ to denote the momentum of the Dirac observable $b^S$.} $P_{b^S}=\frac{8A\mathcal {P}H}{N}\left(\mathcal O_E+\mathcal O_{P_E}\right)$, which is the conjugate momentum to $b^S$, yields
\bq
\label{3a11}
\dot {\mathcal O}_E=-\frac{NP_{b^S}}{2A\mathcal P}-\frac{\delta\hat{C}}{\kappa A\overline C}.
\eq
Moreover, the Hamilton's equation of $P_{b^S}$ can be derived from (\ref{bk2}) and (\ref{bk5}), which takes the form
\bq
\label{3a12}
\dot P_{b^S}=\frac{4H^2\sqrt{A}}{N}b^S+\left(\frac{\dot H}{H}-\frac{\dot N}{N}\right)P_{b^S}.
\eq
Now combining (\ref{3a11}) and (\ref{3a12}), we can obtain the equation of motion for $\mathcal O_E$, namely,
\bq
\label{3a13}
\ddot{\mathcal O}_E+\left(3H-\frac{\dot N}{N}\right)\dot{\mathcal O}_E=\frac{NHb^L}{A}+\frac{N^2}{A}\mathcal O^L_\psi-\left(H-\frac{\dot N}{N}\right)\frac{\delta \hat{C}}{\kappa A\overline C} .
\eq
Finally, the  differential equation which governs the dynamics of the  gauge descriptor $b^L$ can be derived in a straightforward way from its definition (\ref{lg}) and the equations of motion (\ref{bk2}) and (\ref{bk4}). A simple calculation yields,
\bq
\label{3a14}
\dot b^L-\left(\frac{\dot H}{H}-\frac{\dot N}{N}+\frac{3}{2}H+\frac{\kappa N^2 p}{4 H}\right)b^L=N\mathcal O^L_\psi.
\eq

We now compare the results obtained here to the conventional approach where no dust is coupled to gravity. As shown in \cite{Giesel:2018opa}, in the latter case $b^L$ and $\hat{b}^L$ are not gauge invariant but can be understood as the geometrical clocks for the Hamiltonian and spatial diffeomorphism constraints respectively. Hence, these are gauge fixing constraints that are both gauged to zero. Since no dust is present
$\delta  {C}$ and $\delta\hat{C}$ both vanish and this implies $ F^B_\mathrm{dust}$ vanishes as well. Hence, we rediscover the conventional Bardeen equation in this case. If we consider the differential equations in (\ref{3a13}) and (\ref{3a14}) then the limit to the conventional case should be taken with care. While in the dust models $\mathcal{O}_\phi=0$, however in the conventional case this is not true in the longitudinal gauge. For the conventional case in (\ref{3a13}), $\mathcal{O}_E=\dot{\mathcal{O}}_E=\ddot{\mathcal{O}}_E=b^L=\dot{b}^L=\delta \hat{C}=0$ and thus it seems that in this case (\ref{3a13}) forces the Bardeen potential ${\mathcal O}_\psi^L$ to vanish. However, it is important to note  that in the conventional case there is an additional ${\mathcal O}_\phi^L$ term in that differential equation and then (\ref{3a13}) would just yield that the two Bardeen potentials  are not independent but just differ by a sign in our conventions in agreement with the results in \cite{Giesel:2018opa}. The same is true for the differential equation for $b^L$ in (\ref{3a14}) and thus in the conventional case this equations yields no further information.

Finally, we would like to point out that in the conventional two-fluid system consisting of an inflaton field and the dust fields $\left(T, S^i\right)$, there are also six scalar physical degrees of freedom at the linear order of the perturbations. In particular, in addition to the Bardeen potential $\psi^{(gi,L)}$ and its momentum $p^{(gi,L)}_\psi$, one can also construct the other four scalar physical degrees of freedom from the gauge invariant perturbations of the dust fields and their respective momenta. Specifically, in the longitudinal gauge, these gauge invariant perturbations look like
\bqn
\label{dust perturbations l1}
\psi^{(gi,L)}&=&\psi+\frac{4AH^2}{N^2}\left(E+p_E\right)-\frac{1}{3}\Delta E, p^{(gi,L)}_\psi=p_\psi-\left(\frac{AH^2}{N^2}+\frac{\kappa A p}{2}\right)\left(E+p_E\right)+\frac{1}{2}\Delta E+\frac{2}{3}\Delta p_E, \nb\\
\\
\label{dust perturbations l2}
\delta T^{(gi,L)}&=&\delta T+\frac{4AH}{\kappa N}\left(E+p_E\right), \quad \delta P^{(gi,L)}_T=\delta P-\frac{\overline P}{\kappa }\Delta E,\\
\label{dust perturbations l3}
\delta \hat S^{(gi,L)}&=&\delta \hat S,\quad \delta P^{(gi,L)}_{\hat S}=\delta  P_{\hat S}-\frac{\overline P_i}{\kappa}\partial_i E,
\eqn
where $(P, P_i)$ are the conjugate momenta of $(T, S^i)$ and $\overline P$ and $\overline P_i$ denote the background quantities of each variable.
As compared with the physical degrees of freedom in the reduced phase space of the relational formalism with the dust reference clocks, the gauge invariant quantities in the conventional two fluid systems are gauge-invariant only at the linear order of the perturbations.

\subsection{The Mukhanov-Sasaki equation in the Brown-Kucha\v r dust model}
\label{sec:msinbk}
In the Brown-Kucha\v r dust model, the Mukhanov-Sasaki equation can be derived in a similar way as discussed in the above subsection. In the following, we briefly outline the main results and leave the detailed derivations to Appendix \ref{sec:appendix}. The starting point is the  Hamilton's equations of motion for  $\mathcal Q$ and its momentum $P_{\mathcal Q}$ which can be obtained from (\ref{bk3}) and (\ref{bk6}) under the spatially-flat gauge (\ref{sf}),
\bqn
\label{3b2}
\dot{\mathcal Q}&=&\frac{\lambda_\varphi \bar{\pi}_\varphi N}{A^{3/2}}\left(\mathcal O_\phi^S+\frac{P_\mathcal Q}{\bar{\pi}_\varphi}\right),\\
\label{3b3}
\dot{P}_{\mathcal Q}&=&\bar \pi_\varphi \Delta \mathcal O_B^S+\frac{NA^{3/2}}{\lambda_\varphi}\left(-\frac{V_{,\bar \varphi}}{2}\mathcal O_\phi^S+\frac{\Delta \mathcal Q}{A}-\frac{V_{,\bar \varphi \bar \varphi}}{2}\mathcal Q\right).
\eqn
Taking the time derivative of (\ref{3b2}) and then  using (\ref{3b3}), one can obtain a second order differential equation for $\mathcal Q$, which includes terms proportional to $\mathcal O_\phi^S$, $\dot {\mathcal O}_\phi^S$ and $\mathcal O_B^S$ as the source terms. Next, we need to relate these source terms with the gauge invariant perturbations of the dust energy density and momentum density in the spatially flat gauge, which  as discussed in Appendix \ref{sec:appendix} are given respectively by eqs.(\ref{appb5}) and (\ref{deltaEs}).
It turns out that in terms of these gauge invariant perturbations of the dust energy and momentum densities, the Mukhanov-Sasaki equation can be cast into the form
\bqn
\label{MS}
\ddot {\mathcal Q}-\left(\frac{\dot N}{N}-3H\right)\dot{\mathcal Q}&-&\frac{N^2}{A}\Delta \mathcal Q+\left(\frac{N^2}{2}V_{,\bar\varphi\bar\varphi}+\frac{\kappa \bar{\pi}_\varphi N^3 V_{,\bar\varphi}}{2HA^{3/2}}+\frac{\kappa^2 N^4\bar{\pi}^2_\varphi V }{32H^2A^3}\right.\nb \\ &&\left.+\frac{9\lambda_\varphi\kappa N^2 \bar{\pi}^2_\varphi}{8A^3}-\frac{3\lambda^2_\varphi\kappa^2 N^4\bar{\pi}^4_\varphi }{32H^2A^6}\right)\mathcal Q
=F^\mathrm{MS}_\mathrm{dust},
\eqn
with the contributions due to the dust  given explicitly by
\bqn
\label{correction2}
F^{\mathrm{MS}}_\mathrm{dust}&=&- \left(\frac{9\lambda_\varphi \bar \pi_\varphi}{2A^{3/2}}-\frac{3\lambda^2_\varphi \kappa N^2 \bar \pi^3_\varphi }{8H^2 A^{9/2}}+\frac{\kappa N^2 \pi_\varphi V}{8H^2 A^{3/2}}+\frac{N V_{,\bar \varphi}}{H}\right)\frac{\kappa  N^2\overline C b^S }{4A^{3/2}}+\frac{\lambda_\varphi \kappa  \overline C \bar  \pi_\varphi N^2}{4HA^3}\dot b^S \nb\\
&&-\left(\frac{3\lambda^2_\varphi \kappa N^2 \bar\pi^3_\varphi}{8H^2 A^{9/2}}-\frac{9\lambda_\varphi \bar\pi_\varphi}{2A^{3/2}}-\frac{\kappa N^2 \bar \pi_\varphi V}{8H^2A^{3/2}}-\frac{N V_{,\bar\varphi}}{H}\right)\frac{N^2 \delta \hat C}{4A^{3/2}}+\frac{\lambda_\varphi \bar \pi_\varphi N^3}{4HA^3}\delta C
\nb\\
&&-\frac{\lambda_\varphi \kappa  \overline C \bar  \pi_\varphi N^3}{4HA^3}\Delta \mathcal O_E
\eqn
where we have used an intermediate step from Appendix A given in eq.(\ref{appcorrection}).

As a result, we find the Mukhanov-Sasaki variable $\mathcal Q$ and two gauge descriptors $b^S$ and $\hat{b}^S$($=\mathcal{O}_E$)  form a closed system which is governed by Mukhanov-Sasaki equation (\ref{MS}) and the equations for $\mathcal{O}_E$ (\ref{appb9}) and  $b^S$ (\ref{appb8}).
It should be noted that  for the spatially flat case the differential equations for $b^S$ and $\mathcal{O}_E$ are also consistent with the conventional case, in which $b^S=\mathcal{O}_E=0$ and their temporal derivatives vanish as well. In addition $\delta \hat C$ vanishes. Analogous to the longitudinal case, in the conventional case these differential equations would involve $\mathcal{O}^S_\phi$ which gets via these equations related to $\mathcal{Q}$. Using the spatial diffeomorphism constraint in the conventional case relates $\mathcal{O}^S_\phi$ then further to $\mathcal{O}_{p_\psi}$ and $\mathcal{O}_{p_E}$ which agrees exactly with the results obtained in \cite{Giesel:2018opa}. As before the differential equation for $b^S$ and $\mathcal{O}_E$ merge into an identical equation in the limit of the conventional case.

Similar to the case of the longitudinal gauge, in the spatially flat gauge, the conventional two-fluid system consisting of an inflaton field and the dust fields $(T,S^i)$ also contains six physical degrees of freedom in the scalar sector of the linear perturbations, which are the Mukhanov-Sasaki variable $\nu^{(gi,S)}$, the gauge invariant perturbations of the dust fields and their respective momenta. At the linear order in the perturbations, these variables take the form
\bqn
\label{dust perturbations sf1}
\nu^{(gi,S)}&=&\delta \varphi -\frac{\lambda_\varphi \bar \pi_\varphi N}{H A^{3/2}}\left(\psi-\frac{1}{3}\Delta E\right), \quad p^{(gi,S)}_\nu=\delta \pi_\varphi-\bar{\pi}_\varphi \Delta E+\frac{NA^{3/2}V_{,\bar \varphi}}{2\lambda_\varphi H}\left(\psi-\frac{1}{3}\Delta E\right),\nb\\\\
\label{dust perturbations sf2}
\delta T^{(gi,S)}&=&\delta T-\frac{N}{\kappa H}\left(\psi-\frac{1}{3}\Delta E\right), \quad \delta P^{(gi,S)}_T=\delta P-\frac{\overline P}{\kappa }\Delta E,\\
\label{dust perturbations sf3}
\delta \hat S^{(gi,S)}&=&\delta \hat S,\quad \delta P^{(gi,S)}_{\hat S}=\delta  P_{\hat S}-\frac{\overline P_i}{\kappa}\partial_i E.
\eqn
Again, it should be emphasized that the above variables are only gauge invariant at the linear order of the perturbations.

\subsection{The Bardeen  and Mukhanov-Sasaki  equations in the Gaussian dust model}
\label{sec:BardMSinGD}

In the Gaussian dust model,  one can derive the Bardeen and  the Mukhanov-Sasaki equations by  following  the same procedures as in the last two subsections.  The starting point is the equations of motion for the elementary variables which take the similar forms as (\ref{bk1})-(\ref{bk6}), but with a difference that  in the Gaussian dust model, the scalar contribution to the linear perturbations of the shift vector $\mathcal{O}_B$ vanish. Then, one can define the same Bardeen potential and the Mukhanov-Sasaki variable as in (\ref{bpotential}) and (\ref{ms}), respectively.  Following the same strategy as discussed in Sec. \ref{sec:bardeeninbk}, one can obtain the same Hamilton's equations for the Bardeen potential  and its conjugate momentum as in (\ref{3a1}) and (\ref{3a2}), which lead to the same expression of the Bardeen equation as given in (\ref{3a10}), in particular, the dust contributions take exactly the same form as in (\ref{dustinbardeen}).
In addition,  the Hamilton's equation of $P_{b^S}$ takes the same form as in (\ref{3a12}) while the equation for $\mathcal O_E$ in the Gaussian dust model changes to
\bq
\label{3d4}
\dot {\mathcal O}_E=-\frac{NP_{b^S}}{2A\mathcal P} .
\eq
This equation when combined with (\ref{3a12}),  yields
\bq
\label{3d5}
\ddot{\mathcal O}_E+\left(3H-\frac{\dot N}{N}\right)\dot{\mathcal O}_E=\frac{NHb^S}{A},
\eq
or equivalently,
\bq
\label{3d6}
\ddot{\mathcal O}_E+\left(3H-\frac{\dot N}{N}\right)\dot{\mathcal O}_E=\frac{NHb^L}{A}+\frac{N^2}{A}\mathcal O^L_\psi.
\eq
Finally, in order to form a closed system, one also needs the equation of motion for $b^L$ in the Gaussian dust model, which takes the same form as (\ref{3a14}) in the Brown-Kucha\v r dust model.

Similarly, the derivation of the Mukhanov-Sasaki  equation  follows the same strategy in Sec. \ref{sec:msinbk}, which finally leads to the same expressions of the equations as presented in (\ref{MS})  and (\ref{correction2}). Moreover, the equation of motion for $b^S$ also takes the same form as in the Brown-Kucha\v r dust model, which is given by (\ref{appb8}). As a result, in the Gaussian dust model, the Bardeen potential $\mathcal O^L_\psi$, the Mukhanov-Sasaki  variable $\mathcal Q$ and the gauge invariant analogue of the gauge descriptors $b^L$, $b^S$ obey the same equations of motion as their counterparts in the Brown-Kucha\v r dust model. The only difference between the two dust models lies in the  equations of motion of $\mathcal O_E$, which is (\ref{3a13}) in the Brown-Kucha\v r dust model  and (\ref{3d5}) in the Gaussian dust model. Finally, let us briefly comment on the property of the dust model that as discussed in \cite{gt2015} in contrast to the Brown-Kucha\v{r} model the perturbed dust energy density $\delta\mathcal{E}^{\rm dust}$ is not a constant of motion. Since using the gauge invariant analogues of the gauge descriptors as the further two independent gauge invariant variables instead of the gauge invariant analogues of the perturbed energy and momentum density of the dust was justified with $\delta \mathcal E^{\mathrm{dust}}_{\parallelsum}$ and $\delta\mathcal{E}^{\rm dust}$ being constants of motion, the question arises whether we need to modify this choice for the Gaussian dust model.
As shown in \cite{Giesel:2020bht} we have $\mathrm{d} \delta\mathcal{E}^{\rm dust}/\mathrm{d}\tau = \Delta \delta \mathcal E^{\mathrm{dust}}_{\parallelsum}/{A}$. Since for the Gaussian dust model $\delta \mathcal E^{\mathrm{dust}}_{\parallelsum}$ is also a constant of motion we can obtain the time evolution of $\delta\mathcal{E}^{\rm dust}$ if we know $\delta \mathcal E^{\mathrm{dust}}_{\parallelsum}$ and the background evolution. Thus, we can choose the same sets of gauge invariant variables for the Gaussian and Brown-Kucha\v{r} model. \\

\begin{table}[tbh!]\scriptsize
\caption{In the table, we compare the gauge invariant degrees of freedom at the  scalar sector of the linear perturbations in three different models. Both the gauge invariant observables and the equation numbers where they are defined are listed for the spatially flat and the longitudinal gauges. }
\begin{center}
 \begin{tabular}{|p{0.07\linewidth}p{0.05\linewidth}|p{0.08\linewidth}p{0.05\linewidth}||p{0.14\linewidth}|p{0.14\linewidth}||p{0.15\linewidth}p{0.05\linewidth}|p{0.15\linewidth}p{0.05\linewidth}|}
 \hline
 \multicolumn{4}{|p{0.28\linewidth}||}{\centering{The manifestly gauge-invariant perturbation theory with the dust reference clocks}}& \multicolumn{2}{p{0.28\linewidth}||}{\centering{The standard perturbation theory with a single inflaton field}}& \multicolumn{4}{p{0.42\linewidth}|}{\centering{The perturbation theory with a single inflaton and dust with the relational formalism truncated at the linear order with geometric clocks}}\\
 \hline
 \multicolumn{2}{|p{0.12\linewidth}|}{\centering{Spatially flat gauge}}&\multicolumn{2}{p{0.13\linewidth}||}{\centering{Longitudinal gauge}}& \centering{Spatially flat gauge}&\centering{Longitudinal gauge}& \multicolumn{2}{p{0.2\linewidth}|}{\centering{Spatially flat gauge}}&\multicolumn{2}{p{0.2\linewidth}|}{\centering{Longitudinal gauge}}\\
 \hline
 \centering{$\left(\mathcal Q, P_\mathcal{Q}\right)$}& (\ref{ms})&\centering{$\left(\mathcal O^L_\psi,\mathcal O^L_{p_\psi}\right)$}&\centering{(\ref{bardeenp})}&\multirow{3}{9em}{$\left(\nu^{(gi,S)}, p^{(gi,S)}_\nu\right)$}&\multirow{3}{9em}{$\left(\psi^{(gi,L)}, p^{(gi,L)}_\psi\right)$}&$\left(\nu^{(gi,S)}, p^{(gi,S)}_\nu\right)$&(\ref{dust perturbations sf1})&$\left(\psi^{(gi,L)}, p^{(gi,L)}_\psi\right)$&(\ref{dust perturbations l1})\\
 $\left(\hat b^S, P_{\hat b^S}\right)$&(\ref{sf})&$\left(\hat b^L, P_{\hat b^L}\right)$&(\ref{lg})&&&$\left(\delta T^{(gi,S)},\delta P^{(gi,S)}_T\right)$&(\ref{dust perturbations sf2})&$\left(\delta T^{(gi,L)},\delta P^{(gi,L)}_T\right)$&(\ref{dust perturbations l2})\\
 $\left(b^S, P_{b^S}\right)$&(\ref{sf})&$\left(b^L, P_{b^L}\right)$&(\ref{lg})&&&$\left(\delta  \hat S^{(gi,S)},\delta P^{(gi,S)}_{\hat S}\right)$&(\ref{dust perturbations sf3})&$\left(\delta  \hat S^{(gi,L)},\delta P^{(gi,L)}_{\hat S}\right)$&(\ref{dust perturbations l3})\\
 \hline
 \hline
\end{tabular}

\end{center}
\label{t1}
\end{table}

To summarize all the results in the manuscript, we list the main findings in Table \ref{t1}. Here we summarize the main results of gauge invariant variables chosen for the dust model in our work, as compared with the conventional perturbation approach where only a single inflaton field coupled to gravity is considered. This table also presents further insights on these observations by comparing  our work to conventional cosmological perturbation theory based on multi-fluid systems in the background cosmology, see for instance \cite{Hwang:2001fb,Langlois:2008mn,Langlois:2008sg,DeFelice:2009bx,Peter:2015zaa}. But it is to be emphasized that for a consistent comparison one crucially needs to take into account the independent physical degrees of freedom of the various models. In case one follows the conventional approach for the dust models considered here,  a possible choice of gauge invariant variables can be the corresponding gauge invariant versions of the elementary dust clock degrees of freedom $(T, S^i)$ in the scalar sector that are shown in the last column of the table. For a more detailed comparison to multi-fluid systems one needs to consider the specific perfect fluid models under consideration.

\section{Conclusions}
\label{sec:Concl}
\renewcommand{\theequation}{5.\arabic{equation}}\setcounter{equation}{0}
The relational formalism and the reduced phase space approach is a promising avenue to address various conceptual and technical difficulties encountered in canonical
treatments, especially when applied to a quantum gravitational setting. Cosmological perturbation theory provides an  interesting route to test physical implications of the relational formalism.
Since in the conventional approach to cosmological perturbation theory one considers a scalar field coupled to gravity without any reference fields,  the scalar sector of the liner perturbations contains just one degree of freedom. For a given gauge, this degree of freedom
has a natural interpretation for a specific choice of the gauge invariant variable \cite{Giesel:2018opa}. For example, if the chosen gauge is the longitudinal gauge then the
physical degree of freedom is naturally captured by the Bardeen potential, and for the spatially flat gauge this degree of freedom is
encoded in the Mukhanov-Sasaki variable. The number of these physical degrees of freedom increases to three in presence of dust reference fields and the fingerprints of the reference fields influence the dynamical equations for the gauge invariant variables through modifications specific to the choice of reference fields. While one can construct generalizations of the Bardeen potential and the Mukhanov-Sasaki variable in the presence of dust fields,
the additional degrees of freedom create subtleties in comparing these variables to the conventional ones and therefore important gaps existed in investigations of the physical implications of these corrections terms. In previous work the Bardeen-like and Mukhanov-Sasaki-like variables were constructed \cite{ghtw2010II,Giesel:2020bht} and their corresponding dynamical equations were derived which contained extra terms from dust contributions in the relational formalism \cite{ghtw2010II,Giesel:2020bht}. Then one focused on comparing the dynamics of the Bardeen-like and Mukhanov-Sasaki-like variables to the conventional case where only an inflaton coupled to gravity is present. These extra terms, or modifications to the
conventional scenario need to have a physical interpretation but this task turned out to be difficult if one does not take the entire set of physical degrees of freedom into account.
In \cite{Giesel:2020bht} the set of three gauge invariant variables was taken to be the Mukhanov-Sasaki variable together with two elementary Dirac observables in the reduced phase space of the dust models. It turned out that understanding these modifications as functions of this set of gauge invariant variables is quite non-trivial and meanwhile their coupled system of differential equations is also complicated to analyze. Both aspects made a direct comparison to the conventional equations  of the former results a difficult task.
As a consequence, first, a clear physical interpretation of modifications arising from dust degrees of freedom was not available, and second a direct comparison with conventional systems which include an inflaton coupled to gravity, or an inflaton with non-relativistic matter (dust) coupled to gravity which has the same number of degrees of freedom were not available. Accomplishing both of these tasks is important to understand the differences of the relational approach with the conventional approach and to gain insights on the physical meaning of modifications resulting from dust reference fields.  The goal of this manuscript was to explore these issues for the Bardeen potential and Mukhanov-Sasaki variable obtained in the relational formalism using dust reference fields in the Brown-Kuch\v ar and Gaussian dust models and compare with the conventional setting devoid of these reference fields. In particular, our focus was on understanding the additional degrees of freedom tied to the introduction of reference fields through a judicious choice of gauge invariant variables.

 To compare cosmological perturbation theory in the relational framework
based on dust reference fields with the conventional approach, one needed to use suitable gauge invariant variables for the additional degrees of freedom which at the same time simplify the
dynamical system of equations to make them conducive for investigations to compare with other approaches. A pertinent question was also to establish this kind of relationship with a multi-fluid model which has the same number of physical degrees of freedom. This model includes an inflaton coupled with non-relativistic dust matter and gravity. The result of our present analysis was to obtain these sets of gauge invariant variables taking into account the way gauge invariant variables are constructed in the conventional approach for the longitudinal and spatially flat gauge respectively.

To identify these variables for the Brown-Kucha\v{r} and Gaussian dust models, we noted that given a chosen gauge a natural choice of the additional degrees of freedom are the gauge invariant extensions of the energy and momentum density of the dust because the contributions from the dust take a simple form if expressed in terms of them. Using the relationship of the  gauge invariant extensions of the energy and momentum density of the dust with the gauge invariant analogue of the gauge descriptors in the specific gauges, we identified the latter as the gauge invariant variables corresponding to the additional degrees of freedom tied to the dust contributions. However, there is a major difference to the conventional approach here. Since we constructed a map from the independent elementary Dirac observables of the reduced phase space of the dust models to a new set of gauge invariant variables everything is formulated at the manifestly gauge invariant level. This is also the reason why we could identify quantities that usually take the role of gauge descriptors as discussed in \cite{Giesel:2018opa} in the conventional approach without reference fields as gauge invariant variables here.

In our analysis we showed that a transparent physical interpretation of the additional degrees of freedom due to reference fields arises if one uses a different set of variables to express the dynamical equations other than the ones considered earlier. This exercise was carried out for longitudinal and spatially flat gauges and it was shown that in general for each chosen gauge there exists a natural choice for a set of gauge invariant variables. Although, in principle one could use the same set of additional gauge invariant variables for different gauges, the resulting dynamical equations  get unnecessary complicated. This re-enforces the observation, noted earlier for geometrical clocks \cite{Giesel:2018opa}, that even for the case of two-fluid systems like the Brown-Kucha\v{r} and Gaussian dust models, specific gauge choices amount to the choice of a particular set of gauge invariant variables. This result shows that the choice of clock can  have important implications in quantization of these models and their phenomenology in perturbations. Here one would follow the conventional procedure and quantize the dynamics of the gauge invariant degrees of freedom which yields a system of coupled differential equations already in the scalar sector in our case. As far as we can judge from our current analysis none of  the dust reference fields mimicking the geometrical clocks seems to be preferred at this stage. With the techniques introduced in \cite{Pons:2009cz,Pons:2010ad} and applied in \cite{Giesel:2018opa} the gauge invariant variables obtained in the Lagrangian and Hamiltonian framework can be matched. Because one also quantizes the dynamics of the gauge invariant variables in one specific gauge when coming from the Lagrangian formulation, we do not expect any effects on the general covariance of the quantization in the framework presented here which are also not present in the Lagrangian formulation. The additional degrees of freedom that we are forced to quantize compared to the conventional case where only the inflaton as a matter degree of freedom is taken into account, will also be present in multi-fluid systems in the Lagrangian framework. Thus, we should again be able to compare results obtained in the framework here to models in that context and once a quantization has been performed also at the quantum level. To conclude, our manuscript provides an avenue to relate dust reference clocks to conventional methods for inflationary and multi-fluid systems in perturbation theory. Insights gained from our work are expected to be also helpful in understanding the quantization of these systems and associated predictions for cosmological perturbations, especially in canonical quantum gravity.

\section*{Acknowledgements}
This work is  supported by the DFG-NSF grants PHY-1912274 and 425333893.  The authors thank Laura Herold for fruitful discussions at an early stage of this project.

\appendix
\section{Detailed derivations of the Mukhanov-Sasaki equation in the Brown-Kucha\v r dust model}
\renewcommand{\theequation}{A.\arabic{equation}} \setcounter{equation}{0}
\label{sec:appendix}
In this appendix, we derive the Mukhanov-Sasaki equation in the Brown-Kucha\v r dust model in detail.
Taking the time derivative of (\ref{3b2}) and then  using (\ref{3b3}), it is straightforward to arrive at
\bq
\label{app1}
\ddot {\mathcal Q}-\left(\frac{\dot N}{N}-3H\right)\dot{\mathcal Q}-\frac{N^2}{A}\Delta \mathcal Q+\frac{N^2}{2}V_{,\bar \varphi \bar\varphi}\mathcal Q=\frac{\lambda_\varphi N \bar{\pi}_\varphi}{A^{3/2}}\Delta \mathcal O_B^S-N^2V_{,\bar \varphi}\mathcal O_\phi^S+\frac{\lambda_\varphi\bar{\pi}_\varphi N}{A^{3/2}}\dot {\mathcal O}_\phi^S.
\eq
Now, in order to compare the above  Mukhanov-Sasaki  equation with its counterpart in the conventional theory, the first three terms on the right-hand side should be expressed in terms of $\mathcal Q$, $\dot {\mathcal Q}$. Similar to the Bardeen case, one can make use of the perturbed Hamiltonian and diffeomorphism constraints (\ref{dust1})-(\ref{dust2}). First, taking the form of the gauge invariant analogue of the gauge descriptors for the
spatially flat gauge in (\ref{sf}) and the perturbed spatial diffeomorphism constraint in (\ref{dust2}), it is straightforward to show that
\bq
\label{appb4}
\mathcal O_\phi^S=\frac{N\kappa \bar{\pi}_\varphi \mathcal Q}{4 HA^{3/2}}+\frac{N}{4HA^{3/2}}\delta  { \mathcal E}^\mathrm{dust}_{\parallelsum,S},
\eq
here $\delta  { \mathcal E}^\mathrm{dust}_{\parallelsum,S}$ is the gauge invariant perturbation of the dust momentum density in the spatially flat gauge, which  is explicitly related with the geometric and inflaton degrees of freedom via
\bq
\label{appb5}
\delta  { \mathcal E}^\mathrm{dust}_{\parallelsum,S}=-\delta \hat C_{\parallelsum,S}=-\delta  { \hat  C}+\kappa \bar{C}b^S.
\eq
Then, in order to relate the remaining term $\Delta \mathcal O_B^S$ with $\mathcal Q$ and  $\dot {\mathcal Q}$, one should first note that for the spatially flat gauge, we have $\mathcal O_E^S=0$. Considering  the equation of motion  $\dot{\mathcal O}_E^S=0$, which can be directly  read from (\ref{bk2}), we get
\bq
\dot {\mathcal O}_E^S=-4H\left(\mathcal O_E^S+\mathcal O_{p_E}^S\right)+\mathcal O_B^S=-4H\mathcal O_{p_E}^S+\mathcal O_B^S=0.
\eq
On the other hand, according to the definitions in (\ref{newvariables}), it is straightforward to show that  $\mathcal O_\psi^L=\mathcal P^2 \mathcal  O_{p_E}^S$, which relates two different new variables in two different gauges. As a result, we obtain
\bq
\label{appb6}
 \Delta \mathcal O_B^S=\frac{4H}{\mathcal P^2} \Delta \mathcal O_\psi^L.
\eq
Finally, one only needs to make use of (\ref{3a4}) to relate  $ \Delta \mathcal O_\psi^L$ with the  Mukhanov-Sasaki variable and its derivatives. In order to do that, one should first note that
\bq
\frac{V_{,\bar \varphi}}{2\lambda_\varphi}\mathcal O_{\delta \varphi}^L+\frac{\lambda_\varphi \bar \pi_\varphi}{A^3}\mathcal O_{\delta  \pi_\varphi}^L=\frac{V_{,\bar \varphi}}{2\lambda_\varphi}\mathcal Q+\frac{\lambda_\varphi \bar \pi_\varphi}{A^3}P_\mathcal Q.
\eq
Then, (\ref{3a4}) is equivalent to
\bqn
\label{appb7}
-\delta { \mathcal E}^\mathrm{dust}_L=\delta C_L&=&3\mathcal P^2\sqrt{A}\left(\mathcal O_\psi^L-2\mathcal O_{p_\psi}^L\right)-\frac{3\kappa \lambda_\varphi\bar{\pi}^2_\varphi}{A^{3/2}}\mathcal O_\psi^L+4\sqrt A \Delta\mathcal O_\psi^L+\kappa A^{3/2}\left(\frac{V_{,\bar \varphi}}{2\lambda_\varphi}\mathcal Q+\frac{\lambda_\varphi \bar \pi_\varphi}{A^3}P_\mathcal Q\right),\nb\\
&=&\frac{3 H }{N}\left( \kappa \bar{\pi}_\varphi\mathcal Q-\delta { \hat C}_{\parallelsum,L}\right)+4\sqrt A \Delta\mathcal O_\psi^L+\kappa A^{3/2}\left(\frac{V_{,\varphi}}{2\lambda_\varphi}\mathcal Q+\frac{\lambda_\varphi \bar \pi_\varphi}{A^3}P_\mathcal Q\right),
\eqn
where we have used (\ref{3a5}) and the identity
\bq
\mathcal O_{\delta \varphi}^L=\mathcal Q+\frac{\lambda_\varphi \bar{\pi}_\varphi N}{HA^{3/2}}\mathcal O_\psi^L.
\eq
From (\ref{appb7}), one can solve for $ \Delta\mathcal O_\psi^L$ in terms of the Mukhanov-Sasaki  variable $\mathcal Q$, its velocity $\dot {\mathcal Q}$ and also terms involving $\delta {C}_L$ and $\delta  { \hat C}_{\parallelsum,L}$. Hence, this is the last piece we need to rebuild the  Mukhanov-Sasaki  equation in the reduced phase space. Now, combining (\ref{appb4}), (\ref{appb6}) and (\ref{appb7}), it is straightforward to show that the  Mukhanov-Sasaki  equation in the reduced phase space takes the form
\bqn
\label{aMS}
\ddot {\mathcal Q}-\left(\frac{\dot N}{N}-3H\right)\dot{\mathcal Q}&-&\frac{N^2}{A}\Delta \mathcal Q+\left(\frac{N^2}{2}V_{,\bar\varphi\bar\varphi}+\frac{\kappa \bar{\pi}_\varphi N^3 V_{,\bar\varphi}}{2HA^{3/2}}+\frac{\kappa^2 N^4\bar{\pi}^2_\varphi V }{32H^2A^3}\right.\nb \\ &&\left.+\frac{9\lambda_\varphi\kappa N^2 \bar{\pi}^2_\varphi}{8A^3}-\frac{3\lambda^2_\varphi\kappa^2 N^4\bar{\pi}^4_\varphi }{32H^2A^6}\right)\mathcal Q
=F^\mathrm{MS}_\mathrm{dust},
\eqn
with the dust contribution term
\bqn
\label{appcorrection}
F^{\mathrm{MS}}_\mathrm{dust}&=&-\left(\frac{3\kappa \lambda^2_\varphi \bar{\pi}^3_\varphi N^4}{32H^2A^6}-\frac{\kappa \bar{\pi}_\varphi N^4 V}{32 H^2 A^3}-\frac{9\lambda_\varphi \bar{\pi}_\varphi N^2}{8A^3}-\frac{N^3 V_{,\bar \varphi}}{4HA^{3/2}}\right)\delta  { \hat C}_{\parallelsum,S}-\frac{\lambda_\varphi \bar{\pi}_\varphi N^2}{4HA^3}\frac{d}{d\tau}\delta { \hat C}_{\parallelsum,S}\nb\\&&+\frac{\lambda_\varphi \bar{\pi}_\varphi N^3}{4HA^3}\delta C_S .
\eqn
Here we have used (\ref{appb5}) and defined likewise the gauge invariant analogue of the perturbed  momentum density of the dust
\bqn\label{deltaEs}
-\delta { \mathcal E}^\mathrm{dust}_S=\delta C_S&=&\delta C-\kappa  \overline C\left(3\mathcal O_\psi^S+ \Delta \hat b^S\right),\nb\\
&=&\delta C-\kappa \overline C\Delta \mathcal O_E.
\eqn
Similar to the case of the longitudinal gauge we  express $F^{\mathrm{MS}}_\mathrm{dust}$ in terms of the gauge invariant energy and momentum density of the dust and here in addition its temporal derivative. Note, however, that for the spatial flat gauge, these have been obtained using $b^S$ and $\hat{b}^S$. Following the same route as for the Bardeen equation, for spatially flat gauge a convenient choice of the two further gauge invariant variables in addition to the Mukhanov-Sasaki variable $\mathcal{Q}$ is $b^S$ and $\hat{b}^S=\mathcal{O}_E$. Expressed in terms of these set of variables $F^{\mathrm{MS}}_\mathrm{dust}$ takes the following form
\bqn
\label{appcorrection2}
F^{\mathrm{MS}}_\mathrm{dust}&=&- \left(\frac{9\lambda_\varphi \bar \pi_\varphi}{2A^{3/2}}-\frac{3\lambda^2_\varphi \kappa N^2 \bar \pi^3_\varphi }{8H^2 A^{9/2}}+\frac{\kappa N^2 \pi_\varphi V}{8H^2 A^{3/2}}+\frac{N V_{,\bar \varphi}}{H}\right)\frac{\kappa  N^2\overline C b^S }{4A^{3/2}}+\frac{\lambda_\varphi \kappa  \overline C \bar  \pi_\varphi N^2}{4HA^3}\dot b^S \nb\\
&&-\left(\frac{3\lambda^2_\varphi \kappa N^2 \bar\pi^3_\varphi}{8H^2 A^{9/2}}-\frac{9\lambda_\varphi \bar\pi_\varphi}{2A^{3/2}}-\frac{\kappa N^2 \bar \pi_\varphi V}{8H^2A^{3/2}}-\frac{N V_{,\bar\varphi}}{H}\right)\frac{N^2 \delta \hat C}{4A^{3/2}}+\frac{\lambda_\varphi \bar \pi_\varphi N^3}{4HA^3}\delta C
\nb\\
&&-\frac{\lambda_\varphi \kappa  \overline C \bar  \pi_\varphi N^3}{4HA^3}\Delta \mathcal O_E.
\eqn
To obtain the coupled system of differential equations for the set ($\mathcal{Q}$, $b^S$,$\mathcal{O}_E$), we need to consider the equations of motion for $\mathcal O_E$ and $b^S$. Similar to the Bardeen case, the equation of motion for $\mathcal O_E$ takes the  form
\bq
\label{appb9}
\ddot{\mathcal O}_E+\left(3H-\frac{\dot N}{N}\right)\dot{\mathcal O}_E=\frac{NHb^S}{A}-\left(H-\frac{\dot N}{N}\right)\frac{\delta \hat C}{\kappa A\overline C},
\eq
which is equivalent to (\ref{3a13}) as $b^S=N\mathcal O^L_\psi/H+b^L$. Finally,  the  equation of motion for $b^S$ can be derived from its definition (\ref{sf}) and the equations of motion (\ref{bk1})-(\ref{bk2}) and (\ref{bk4})-(\ref{bk5}), which turns out to be
\bq
\label{appb8}
\dot b^S-\left(\frac{\dot N}{N}-\frac{\dot H}{H}-\frac{\kappa \lambda_\varphi N^2 \bar\pi^2_\varphi}{4HA^3}\right)b^S=\frac{N^2}{4A^{3/2}H}\Bigg\{\delta \hat C -\kappa \bar\pi_\varphi Q \Bigg\}.
\eq


\end{document}